\documentclass[amsmath,prd,floatfix,preprintnumbers,nofootinbib]{revtex4}
\usepackage{graphicx} \allowdisplaybreaks[4]

\newcommand{\ttwo}[0]{\mbox{$\frac{t}{2}$}}
\newcommand{\pslash}[0]{p\hspace*{-1.7mm}/}

\begin{document}

\preprint{NT@UW 04-021}

\title{Flavour singlet physics in lattice QCD with background fields}

\author{W. Detmold} \affiliation{Department of Physics, University of
  Washington, Box 351560, Seattle, WA 98195, U.S.A.}

\begin{abstract}
  We show that hadronic matrix elements can be extracted from lattice
  simulations with background fields that arise from operator
  exponentiation. Importantly, flavour-singlet matrix elements can be
  evaluated without requiring the computation of disconnected
  diagrams, thus facilitating a calculation of the quark contribution
  to the spin of the proton and the singlet axial coupling, $g_A^0$.
  In the two nucleon sector, a background field approach will allow
  calculation of the magnetic and quadrupole moments of the deuteron
  and an investigation of the EMC effect directly from lattice QCD.
  Matrix elements between states of differing momenta are also
  analysed in the presence of background fields.
\end{abstract}


\maketitle

The composition of the spin of the proton is a question of
long-standing interest since experimentally it is known that only
about 20\% of the total spin of the proton is carried by the quark
helicity \cite{spin}. Contributions from quark orbital angular
momentum and from gluon angular momentum are obviously important, but
are currently unknown. Recently Ji \cite{Ji:1996ek} has shown that a
well-defined decomposition of the proton spin exists,
\begin{equation} 
  \label{eq:6}
  \frac{1}{2}=J_q+J_g=\frac{1}{2}\Sigma_q+L_q+J_g\,,
\end{equation}
where each term is gauge invariant (though scheme and scale dependent)
and given by proton matrix elements of the operators
\begin{eqnarray}
  \label{eq:7}
  {\bf J}_q&=&\int d^3{\bf r}\,{\bf r}\times {\bf T}_q \\
  &=& 
  \frac{1}{2}\mbox{\boldmath$\Sigma$}_q+{\bf L}_q \equiv \int d^3{\bf
    r}\,  
  \overline\psi \left[\frac{1}{2}\mbox{\boldmath$\gamma$}\gamma_5+\gamma_0{\bf
      r}\times(-i{\bf D})\right]\psi\,, 
  \nonumber \\
  {\bf J}_g&=&\int d^3{\bf r}\,{\bf r}\times ({\bf E}\times{\bf B})\,,
\end{eqnarray}
where colour indices are suppressed, ${\bf T}_q^j=T_q^{j0}$ and
$T_q^{\mu\nu}=\overline{\psi}\gamma^{\{\mu}i\tensor{D}^{\nu\}}\psi$ is
the traceless, symmetric part of the quark energy-momentum tensor
($\tensor{D}=\frac{1}{2}[\roarrow{D}-\loarrow{D}]$).  The quark
helicity contribution, $\Sigma_q$, has been measured in polarised
deep-inelastic scattering, but the separation of the quark orbital
angular momentum and the total gluon angular momentum has not yet been
determined from experiment though it can be extracted from
deeply-virtual Compton scattering.

\begin{figure}[!t]
  \centering \includegraphics[width=8cm]{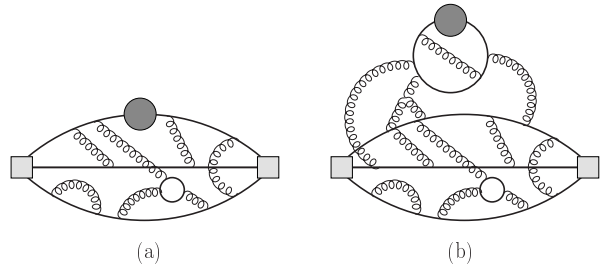}
  \caption{Connected (a) and quark-line disconnected (b)
    contributions to a baryon matrix element. The squares correspond
    to the baryon source and sink and the dark circle to the operator
    insertion.}
  \label{fig:conn}
\end{figure}

A reliable prediction of the various contributions to Eq.~(\ref{eq:6})
from QCD would obviously be an important achievement.  However, in QCD
this question is non-perturbative since it involves hadronic-scale
physics, and to address the issue from first principles, one must use
lattice QCD. Indeed, proton matrix elements of both the helicity
operator, $\overline{\psi}\gamma^\mu\gamma_5\psi$,
\cite{lattice_gA0,lattice_helicity} and the energy momentum tensor,
$T^{\mu\nu}_q$, \cite{Mathur:1999uf, Gadiyak:2001fe, Hagler:2003jd,
  Gockeler:2003jf} have been computed and some estimates of the
partitioning of the proton spin have been made.  These calculations
have been performed at unphysically large quark masses and on modest
volumes and such parametric limitations make the connection of these
results to the physical world non-trivial, though progress continues
to be made in this regard
\cite{ChPT,Chen:2001pv,Belitsky:2002jp,FVGA,FVTTMM}. Such issues must
be confronted for all hadronic quantities extracted from lattice
simulations.  However, the total quark helicity (singlet axial
coupling) and angular momentum content of the proton are flavour
singlet quantities ($g_A^0 \sim \Sigma_q\sim \Sigma_u+\Sigma_d$ and
$J_q\sim J_u+J_d$) and lattice calculations of them are further
plagued by so-called quark-line disconnected contributions in which
the relevant operator is inserted on a quark line connected to the
proton source only by gluons as in Fig.~\ref{fig:conn}(b). These terms
are notoriously difficult to compute (see Ref.~\cite{Gusken:1999te}
for a review) and result in relative errors an order of magnitude
larger than the corresponding connected contributions.

Numerous other quantities such as the strangeness magnetic moment and
the pion-nucleon sigma term suffer from the same difficulties. In the
two nucleon sector, notable examples are the deuteron magnetic and
quadrupole moments, and the deuteron structure function, $F_2^d(x)$,
(accessible in lattice QCD only through two-particle matrix elements
of twist-two operators \cite{EMCpaper}) that gives the simplest
manifestation of the EMC effect \cite{Norton:2003cb}. Even were it
feasible to calculate five point correlators on present day computers,
these quantities would again require orders of magnitude more
computational effort than their flavour non-singlet analogues.

In this article, we highlight an alternative method to calculate
operator matrix elements that eliminates the issue of disconnected
contributions at the expense of requiring the generation of additional
ensembles of gauge configurations. This essentially involves computing
two-point correlators in the presence of generalised background fields
which we show arise from exponentiating the operator whose matrix
element we wish to calculate into the QCD action (see
Ref.~\cite{Fucito:ff} for an early example).  Since disconnected
contributions are included automatically, this approach may provide a
means of extracting reliable results for singlet matrix elements.

Electroweak background field methods have a long history in lattice
QCD. In the early 1980s, Martinelli {\it et al.}
\cite{Martinelli:1982cb} and Bernard {\it et al.}
\cite{Bernard:1982yu} performed the first calculations of the magnetic
moments of the proton, neutron and $\Delta$-resonance by calculating
the baryon masses on a lattice immersed in a constant magnetic field.
This approach has since been extended to calculate electric and
magnetic polarisabilities of various hadrons
\cite{Fiebig:1988en,Christensen:2002wh, Zhou:2002km, Burkardt:1996vb}.
Fucito {\it et al.} \cite{Fucito:ff} also showed that a background
axial field could be used to determine the isovector axial coupling of
the nucleon, $g_A$.  Recently it has been demonstrated
\cite{Detmold:2004qn} that background electroweak fields can be used
to probe the electroweak properties of two-nucleon systems.
Measurement of two particle energy levels at finite volume in magnetic
and weak background fields is sufficient to determine the weak- and
magnetic- moments of the deuteron, and the threshold cross-sections
for radiative capture ($n\, p\to d \gamma$) and weak-disintegration of
the deuteron (e.g. $\nu_x d\to \nu_x n\, p$ $[x=e,\mu,\tau]$ and
$\overline{\nu}_e d\to e^+ n\, n$).

In this work, we extend these analyses to consider more general
background fields that correspond to exponentiation of a larger class
of operators. We additionally consider the use of background fields to
extract operator matrix elements between states of differing momenta.
Such matrix elements determine the electromagnetic form-factors and
moments of generalised parton distributions.  Whilst much of the
following analysis will be equally applicable in the pure-gauge and
mesonic sectors, we restrict our discussion to baryons.

{\it One baryon:} In the single baryon sector, one is interested in
calculating matrix elements of the form
\begin{equation}
  \label{eq:2}
  \langle B^\prime({\bf p}^\prime,s^\prime) | {\cal O} |
  B({\bf p},s)\rangle \sim 
\overline{u}_{B^\prime}({\bf p}^\prime,s^\prime) {\cal M}_{B^\prime B}[{\cal O}]
u_B({\bf p},s)\,, 
\end{equation}
where $q=p^\prime-p$ is the relative momentum of the baryons $B$ and
$B^\prime$ ($s$ and $s^\prime$ label their spin) and $u_B({\bf p},s)$
is a baryon spinor. The Dirac structure ${\cal M_{B^\prime B}[{\cal
    O}]}$ is parameterised by a set of scalar form factors,
$F_{i}(q^2)$, depending on the particular operator, ${\cal O}$, and
the external states (for the case of the energy-momentum tensor, we
shall be more specific below).  Typically, such form factors are
calculated by computing ratios of two and three point correlators.  We
define
\begin{eqnarray}
  \label{eq:3}
  G_{BB}({\bf p},t;\Gamma) = 
  \sum_{\bf x} e^{-i{\bf p}\cdot{\bf x}} \Gamma_{\beta\alpha}
  \langle  0|\chi^{(B)}_\alpha({\bf x},t) \overline{\chi}^{(B)}_{\beta}
  (0,0)|0\rangle \,, 
\end{eqnarray}
and
\begin{eqnarray}
  \label{eq:55}
  G_{B^\prime{\cal O}B}({\bf p},{\bf p^\prime},t,\tau;\Gamma^\prime) &= & \sum_{{\bf x},{\bf y},{\bf z}}
  \Gamma^\prime_{\beta\alpha} e^{-i{\bf p}\cdot{\bf x}}
  e^{i{\bf p^\prime}\cdot{\bf z}} 
  \langle 0|\chi^{(B^\prime)}_\alpha({\bf x},t){\cal O}({\bf y},\tau)
  \overline{\chi}^{(B)}_{\beta}({\bf z},0)|0\rangle \,,
\end{eqnarray}
where $\Gamma$ and $\Gamma^\prime$ are Dirac projectors (the exact
form of which depends on the operator under consideration), and
$\chi_\alpha^{(B)}$ ($\overline{\chi}_{\beta}^{(B)}$) is an
interpolating field for the baryon sink (source); for the proton, a
common choice is $\chi^{(p)}_\alpha(x)=\epsilon_{ijk}u^i_\alpha(x)
u^j_\beta(x)(C\gamma_5)_{\beta\gamma} d^k_\gamma(x)$. It is easy to
show (by inserting complete sets of states, see e.g.
Ref~\cite{Hagler:2003jd}) that the ratio
\begin{eqnarray}
  \label{eq:4}
  R({\bf p^\prime},{\bf p},t_{\rm sink},\tau;\Gamma,\Gamma^\prime)=
  \frac{G_{B{\cal O}B}({\bf p^\prime},{\bf p},t_{\rm sink},\tau;\Gamma^\prime)}
  {G_{BB}({\bf p^\prime},t_{\rm sink};\Gamma)}
\,\left[\frac{G_{BB}({\bf p^\prime},\tau;\Gamma) G_{BB}({\bf
        p^\prime},t_{\rm sink};\Gamma) G_{BB}({\bf p},t_{\rm sink}-\tau;\Gamma)} 
    {G_{BB}({\bf p},\tau;\Gamma) G_{BB}({\bf p},t_{\rm sink};\Gamma) G_{BB}({\bf
        p^\prime},t_{\rm sink}-\tau;\Gamma)}\right]^{1/2}\,, 
\end{eqnarray}
will exhibit a plateau over a range of time slices $a\ll\tau\ll t_{\rm
  sink}$ where the sink time, $t_{\rm sink}$, is held fixed. For time
slices within this plateau and appropriate choices of the Dirac
projectors, this ratio is proportional to the form-factors under
consideration. This approach and its variants (see {\it e.g.}
Refs.~\cite{Martinelli:1987bh,Leinweber:1990dv}) are referred to as
the operator insertion method. For flavour singlet operators,
calculation of $G_{B{\cal O}B}$ requires evaluation of both types of
diagram in Fig.~\ref{fig:conn}.

For a local operator, ${\cal O}(z)$, at most bilinear in quark fields,
it is possible to calculate the same matrix element by measuring a
two-point correlator on an ensemble of field configurations generated
with a modified Boltzmann weight. To see this, we add a term
\begin{eqnarray}
  \label{eq:11}
  \int d^d z \, {\cal O}(z)\cdot \Omega(z)\,,
\end{eqnarray}
to the usual QCD action, $S_{\rm QCD}$, with a fixed external source
field $\Omega(z)$. If ${\cal O}(z)$, which depends on the quark
($\psi$) and gluon ($A_\mu$) fields and their derivatives, carries
Lorentz or flavour indices, so does $\Omega(z)$.  In the theory
described by this action, the Euclidean space correlator of two baryon
interpolating fields is given by:
\begin{eqnarray}
  \label{eq:1}
  \langle 0| {\chi}_\alpha(x) \overline\chi_{\alpha^\prime}(y) |0
  \rangle_\Omega &=& 
  \frac{1}{{\cal Z}[\Omega]}\int {\cal D}A_\mu  {\cal D}\psi {\cal
    D}\overline{\psi}  {\chi}_\alpha(x)\overline\chi_{\alpha^\prime}(y)
  \exp\left(- S_{\rm QCD}[A,\psi,\overline{\psi}]  - \int d^dz\; {\cal
      O}(z) \cdot \Omega(z)\right)   
  \\
 \label{eq:111}
   &\hspace*{-3cm}\xrightarrow{|\Omega|\to0}&\hspace*{-1cm}
  \frac{1}{{\cal Z}[0]}\int {\cal D}A_\mu {\cal D}\psi
  {\cal D}\overline{\psi} {\chi}_\alpha(x)\overline\chi_{\alpha^\prime}(y) 
  e^{- S_{\rm QCD}}\left[1+\frac{1}{{\cal Z}[0]}\int d^dz\Omega(z) \int
    {\cal D}A_\mu {\cal D}\psi 
    {\cal D}\overline{\psi}{\cal O}(z)e^{-S_{\rm QCD}}\right]
  \nonumber \\
  &&\hspace*{3cm}
-  \frac{1}{{\cal Z}[0]}\int d^dz \, \Omega(z)\int {\cal D}A_\mu {\cal D}\psi
  {\cal D}\overline{\psi}  {\chi}_\alpha(x) {\cal O}(z) 
  \overline\chi_{\alpha^\prime}(y)e^{-  S_{\rm QCD}} 
  +{\cal O}(|\Omega|^2)
  \nonumber \\
  &\hspace*{-3cm}=&\hspace*{-1cm} 
  \langle 0|{\chi}_\alpha(x)\overline\chi_{\alpha^\prime}(y)|0\rangle_0 
  \left[1+\int d^dz\, \Omega(z) \;\langle 0|{\cal O}(z)  |0
    \rangle_0\right] - \int d^dz\, \Omega(z) \;\langle 0|
  {\chi}_\alpha(x) {\cal O}(z) \overline\chi_{\alpha^\prime}(y) |0
  \rangle_0 +{\cal O}(|\Omega|^2)\,,
  \nonumber
\end{eqnarray}
where ${\cal Z}[\Omega]$ is the partition function of the modified
theory, $\langle\cdots\rangle_\Omega$ indicates a correlator evaluated
in this theory, and $|\Omega|\to0$ is a shorthand for requiring all
tensor components of $\Omega(z)$ to be small compared to $\Lambda_{\rm
  QCD}^{{\rm dim}\,\Omega}$ (from Eq.~(\ref{eq:11}), ${\rm
  dim}\,\Omega=d-{\rm dim}\,{\cal O}$).  For operators for which
$\langle 0|{\cal O}(z) |0 \rangle=0$ [a notable exception being
$\overline\psi(z)\psi(z)$], the (completely) disconnected contribution
vanishes and we see that to leading order in the strength of the
background field, $\Omega(z)$, the difference between the two-point
correlators with $\Omega(z)=0$ and $\Omega(z)\ne0$ is proportional to
the form factors that we wish to extract. For example, defining the
external field generalisation of Eq.~(\ref{eq:3}) as
\begin{equation}
G_{B^\prime B}^\Omega({\bf p}^\prime,{\bf p},t;\Gamma)=
\sum_{{\bf x},{\bf y}} \Gamma_{\beta\alpha} e^{-i{\bf
    p}^\prime\cdot{\bf x}} e^{i{\bf p}\cdot{\bf y}} 
\langle 0 | \chi^{(B^\prime)}_\alpha({\bf x},t) \overline{\chi}^{(B)}_{\beta}
  ({\bf y},0)|0\rangle_\Omega \,,
\end{equation}
then
\begin{eqnarray}
\label{eq:ME_analysis}
\frac{G_{BB}({\bf p},t;\Gamma)- G_{BB}^\Omega({\bf p},{\bf p},t;\Gamma)}
{G_{BB}({\bf p},t;\Gamma)}  
&\xrightarrow[|\Omega|\to0]{t\to\infty}&\frac{\widetilde\Omega(0)}{2\, E_B({\bf p})} 
\frac{{\rm tr}\left[\Gamma \left(-i\pslash+M_B \right) {\cal M}_{BB}[{\cal
      O}] \left(-i\pslash+M_B \right)\right]} {{\rm tr}\left[\Gamma
    \left(-i\pslash+M_B \right)\right]} \,,
\end{eqnarray}
where $\widetilde{\Omega}(q)$ is the (discrete) Fourier transform of
$\Omega(z)$, $M_B$ is the ground state baryon mass, $E_B({\bf
  p})=\sqrt{M_B^2+|{\bf p}|^2}$ and terms of ${\cal O}(|\Omega|^2)$
are ignored.  With appropriate choices of $\Gamma$, this gives the
required form-factors at $q^2=0$. The non-forward case is discussed
below.

If ${\cal O}(z)$ depends on quark fields, the effects of the external
field will manifest themselves in the quark determinant\footnote{One
  must be careful that the additional term in the action does not
  destroy the positivity of the determinant and thereby the
  probabilistic nature of the gauge integration measure. For example,
  adding the term $i\mu\int d^dz\, \overline\psi(z)\gamma_0\psi(z)$
  would be problematic.} and valence quark propagators after the
fermionic functional integration is performed in Eq.~(\ref{eq:1}). In
a lattice simulation where one approximates the integral over the
field configurations by importance sampling, each choice of background
field requires an additional ensemble of appropriately weighted gauge
field configurations to be generated.  We note that in quenched QCD a
given set of gauge configurations can be modified to incorporate the
effects of the exponentiated operator if it is composed of purely
quark fields (such as the local electromagnetic and axial currents) as
the operator and gauge field decouple in the absence of vacuum
polarisation by dynamical sea quark loops. In more general cases, new
gauge configurations are required even in the quenched theory.

This exponentiated operator (external field) method is exactly what
has been used to calculate the quenched magnetic moment of the proton
in Refs.~\cite{Martinelli:1982cb,Bernard:1982yu}. Here the addition to
the Lagrangian is given by
\begin{equation}
  \label{eq:33}
  {\cal O}^\mu_{EM}(z) = \overline{\psi}(z)\gamma^\mu\psi(z),
  \quad
  \Omega_\mu(z) = \frac{e\, B}{2}
  {\tiny\begin{pmatrix}-z_2\\z_1\\0\\0\end{pmatrix}}\,, 
\end{equation}
producing a constant magnetic field in the $z_3$ direction over all
lattice sites (ignore issues of periodicity at finite volume). In a
lattice regularisation, the simplest local transcription of this
current is not conserved and a multiplicative renormalisation factor
is required; alternatively the lattice version of the conserved
current \cite{Karsten:1980wd,Martinelli:1987bh} can be used. For small
field strengths, $B$, differences
\cite{Martinelli:1982cb,Bernard:1982yu} between correlators of spin-up
and spin-down baryons then determine the relevant magnetic moment,
$\mu_B$, since the lowest energy eigenstates behave as $M_B^{\pm}=M_B
\pm \mu_B |B| +{\cal O}(|B|^2)$ where $\pm$ refers to spin
anti-aligned or aligned with the magnetic field. To extract the
magnetic moment, one requires one set of pure QCD gauge configurations
generated with $B=0$ to determine the mass, $M_B$, and another
independent set of configurations generated with the modified action
with $a|e\,B|\ll\Lambda_{\rm QCD}$ to determine the shifted masses,
$M_B^\pm$. In practice, a few ensembles of configurations generated
with different values of $B\ne0$ may be needed to uniquely determine
the mass shift linear in $B$.

The quark angular momentum contribution to the proton spin can be
measured from a forward matrix element \cite{Gadiyak:2001fe} in a
background field since it can be defined as a spatial moment, ${\bf
  J}_q=\int d^3{\bf r}\; {\bf r}\times {\bf T}_q$ (as with the
magnetic moment, $\mbox{\boldmath$\mu$}=\frac{1}{2}\int d^3{\bf
  r}\;{\bf r}\times{\bf j}^{EM}$ \cite{Jackson}); in the background
field approach\footnote{The difficulties of using continuum moment
  equations on the lattice highlighted in Ref.~\cite{Wilcox:2002zt} do
  not apply to background field calculations.}, one would add $\Omega$
times this integral to the QCD action and look at the shift in the
exponential fall off of the two-point correlator exactly as for the
magnetic moment. Since twist-two operators such as the energy-momentum
tensor are not conserved even in the continuum, there is a
multiplicative renormalisation factor that must be computed to obtain
a final result.  The renormalisation factors for various twist-two
operators have been calculated \cite{Capitani:1994qn}.  A more generic
approach not relying on moment equations (and hence valid for more
general operators that cannot be written in such a form) can also be
used to determine $J_q$.  Nucleon matrix elements of the helicity
independent, dimension four, twist-two quark operator take the form
\begin{eqnarray}
  \langle N({\bf p}^\prime,s^\prime)|\overline\psi_f \gamma^{\{\mu} i
  \tensor{D}^{\nu\}}\psi_f |N({\bf p},s)\rangle = 
  \overline{u}_N({\bf p}^\prime,s^\prime)\Big[\gamma^{\{\mu}  \overline{p}^{\nu\}}
  A_f(q^2) 
  - i \frac{q_\rho\sigma^{\rho\{\mu}}{2M}
  \overline{p}^{\nu\}} B_f(q^2) 
  +\frac{q^{\{\mu}q^{\nu\}} }{M} C_f(q^2)\Big]u_N({\bf p},s) \,,
  \label{eq:8}
\end{eqnarray}
where $\overline{p}=\frac{1}{2}\left(p+p^\prime\right)$ and
$\{\ldots\}$ indicates symmetrisation of indices and subtraction of
traces.  In terms of the form-factors, $A_f$, $B_f$ and $C_f$, the
total quark angular momentum content for flavour $f$ is then given by
\cite{Ji:1996ek} $J_f = \frac{1}{2}\left[A_f(0) + B_f(0)\right]$.
Since $B_f(q^2)$ is always accompanied by a factor of the momentum
transfer, the twist-two matrix elements must be calculated at $q\ne0$
(using Eq.~(\ref{eq:4}) for example) and then extrapolated to the
forward limit.  This necessarily leads to some uncertainty as the
minimum available non-zero lattice momentum, $q_{\rm min}=2\pi/L$, is
set by the lattice size, $L$. For current lattice simulations, $L\alt
3$~fm, so $q_{\rm min}\agt 0.4$~GeV. Given such a large momentum
extrapolation, the use of the spatial moment definition in
Eq.~(\ref{eq:7}) within the background field approach may provide the
cleanest\footnote{Another possibility is to consider the correlator
  subject to twisted boundary conditions \cite{Bedaque:2004kc} (which
  correspond to a particular choice of external vector field) as they
  can reduce the minimum available momentum.}  determination of $J_q$.

Exponentiated operator methods can be used to calculate such
off-forward matrix elements. To see this, we consider the following
correlator (${\bf p}^\prime\ne{\bf p}$) in the presence of an
exponentiated operator $\int d^dz \,\Omega(z) \cdot {\cal O}(z)$:
\begin{widetext}
  \begin{eqnarray}
    \label{eq:9}
    G_{B^\prime B}^\Omega({\bf p}^\prime,{\bf p},t;\Gamma) &=&
\sum_{{\bf x},{\bf y}} \Gamma_{\beta\alpha} e^{-i{\bf
    p}^\prime\cdot{\bf x}} e^{i{\bf p}\cdot{\bf y}} 
\langle 0 | \chi^{(B^\prime)}_\alpha({\bf x},\ttwo) \overline{\chi}^{(B)}_{\beta}
  ({\bf y},-\ttwo)|0\rangle_\Omega 
\\
&=&-\sum_{{\bf x},{\bf y}} \sum_{{\bf z},\tau}
\Gamma_{\beta\alpha}\Omega({\bf z},\tau)  e^{-i{\bf
    p}^\prime\cdot{\bf x}} e^{i{\bf p}\cdot{\bf y}} 
\langle 0 | \chi^{(B^\prime)}_\alpha({\bf x},\ttwo) {\cal O}({\bf z},\tau)\overline{\chi}^{(B)}_{\beta}
  ({\bf y},-\ttwo)|0\rangle
    +{\cal O}(|\Omega|^2)
\nonumber
\\&=&
-\sum_{n,{\bf p}_n,s^\prime}\sum_{m,{\bf p}_m,s}\sum_{{\bf x},{\bf y}}
    \sum_{{\bf z},\tau} \Gamma_{\beta\alpha}  \Omega({\bf z},\tau) 
e^{-i\left({\bf p}_n -{\bf p}^\prime\right)\cdot{\bf x}} 
e^{i\left({\bf p}-{\bf p}_m\right)\cdot{\bf y}} 
    e^{-i({\bf p}_m-{\bf p}_n)\cdot{\bf z}} 
    e^{-(E_n+E_m)t/2}e^{(E_n-E_m)\tau}
\nonumber \\
&&\hspace*{2cm}\times
\langle 0 | \chi^{(B^\prime)}_\alpha(0)|n,{\bf
  p}_n,s^\prime\rangle 
\langle n,{\bf p}_n,s^\prime |{\cal O}(0) | m,{\bf p}_m,s \rangle
\langle m,{\bf p}_m,s  | \overline{\chi}^{(B)}_{\beta}
  (0)|0\rangle
    +{\cal O}(|\Omega|^2)
\nonumber
\\&=&
-\sum_{n,m}    e^{-(E_n+E_m)t/2} \Gamma_{\beta\alpha}
\sum_{s,s^\prime} 
    \sum_{{\bf z},\tau} 
    e^{-i({\bf p}-{\bf p}^\prime)\cdot{\bf z}} 
    e^{(E_n-E_m)\tau}  \Omega({\bf z},\tau) 
\nonumber \\
&&\hspace*{2cm}\times
\langle 0 | \chi^{(B^\prime)}_\alpha(0)|n,{\bf
  p}^\prime,s^\prime\rangle 
\langle n,{\bf p}^\prime,s^\prime |{\cal O}(0) | m,{\bf p},s \rangle
\langle m,{\bf p},s  | \overline{\chi}^{(B)}_{\beta}
  (0)|0\rangle
    +{\cal O}(|\Omega|^2)
\nonumber \\
&\xrightarrow[|\Omega|\to0]{t\to\infty}&
-e^{-\left[E_{B^\prime}+E_B\right]t/2}\sum_{s,s^\prime}  \Gamma_{\beta\alpha}
     \widetilde\Omega(q) 
\nonumber \\
&&\hspace*{2cm}\times
\langle 0 | \chi^{(B^\prime)}_\alpha(0)|B^\prime({\bf
  p}^\prime,s^\prime)\rangle 
\langle B^\prime({\bf p}^\prime,s^\prime) |{\cal O}(0) | B({\bf p},s) \rangle
\langle B({\bf p},s)  | \overline{\chi}^{(B)}_{\beta}
  (0)|0\rangle \,,
\nonumber
\end{eqnarray}
where $\widetilde{\Omega}(q)$ is again the (discrete) Fourier
transform of $\Omega(z)$, $E_B$ and $E_{B^\prime}$ are the energies of
the lowest $B$ and $B^\prime$ states with momentum ${\bf p}$ and ${\bf
  p}^\prime$.  Since energy and momentum are conserved in pure QCD
simulations, $G_{B^\prime B}^{0}({\bf p}^\prime, {\bf
  p},t;\Gamma)\sim\delta_{B^\prime B}\delta^3({\bf p}^\prime-{\bf
  p})$. However, by including an external source that is inhomogeneous
[$\Omega=\Omega({\bf z},\tau)$] in the action, energy and/or momentum
can be injected through the operator coupled to the source and
off-forward correlators can have non-zero values. By choosing
appropriate external fields, the form factors implicit in ${\cal
  M}_{B^\prime B}[{\cal O}]$ are easily determined from
\begin{eqnarray}
  \label{eq:10}
  \frac{G_{B^\prime B}^{\Omega}({\bf p}^\prime, {\bf
  p},t;\Gamma^\prime)}{\sqrt{G_{B^\prime B^\prime}^{0}({\bf p}^\prime, {\bf
  p}^\prime,t;\Gamma) \, G_{B B}^{0}({\bf p}, {\bf p},t;\Gamma)}}
  = -\frac{\widetilde{\Omega}(q)}{2\sqrt{M_B M_{B^\prime}}}
  \frac{{\rm tr}\left[\Gamma^\prime(-i\pslash^\prime+M_{B^\prime}){\cal
        M}_{B^\prime B}[{\cal O}](-i\pslash+M_{B})\right]}{\sqrt{{\rm
        tr}\left[\Gamma(-i\pslash^\prime+M_{B^\prime})\right] \, {\rm
        tr}\left[\Gamma(-i\pslash+M_{B})\right]}}    
  +{\cal O}(|\Omega|^2)\,, 
\end{eqnarray}
\end{widetext}
with relevant choices of $\Gamma$ and $\Gamma^\prime$. The simplest
choice would be an external field that is a plane wave, $\Omega(z)
\sim \exp(-i\, \hat{q}\cdot z)$, such that $\widetilde{\Omega}(q)\sim
\delta^4(q-\hat{q})$, but other choices are possible. Correlators in
which both the source and sink momentum are fixed are easily computed
\cite{Gockeler:1998ye}.

This method will be applicable not only for off-forward matrix
elements of the electromagnetic current (giving the Dirac and Pauli
form factors) and the energy-momentum tensor (giving the form factors
in Eq.~(\ref{eq:8})), but also for other quark bilinear operators.
Important cases are the towers of twist-two operators ${\cal
  O}_{\Gamma,f}^{\mu_0\ldots\mu_n} = \overline\psi_f \Gamma^{\{\mu_0}
i\tensor{D}^{\mu_1} \ldots i\tensor{D}^{\mu_n\}} \psi_f$ where
$\Gamma^\mu=\gamma^\mu,\, \gamma^\mu\gamma_5, \, \sigma^{\alpha\mu}$
and $\{\ldots\}$ again indicates symmetrisation of indices and
subtraction of traces. The non-forward matrix elements of these
operators correspond to moments of generalised parton distributions
(see Ref.~\cite{Diehl:2003ny} for a recent review).  Finally we note
that hadronic matrix elements of purely gluonic operators would also
be accessible through the operator exponentiation approach but
operators involving more than two quark fields cannot be exponentiated
as they would prevent the fermionic functional integrals from being
integrated exactly.

For many quantities the background field approach would not be
particularly appealing as the added overhead (with respect to
calculating a baryon two point correlator) of computing the quark
propagators required in the standard three point correlator analysis
is minimal as compared to the cost of generating additional ensembles
of dynamical gauge configurations; particular examples are the moments
of the isovector parton distribution functions.  However, for flavour
singlet quantities the situation may be reversed. In the operator
insertion approach, one must evaluate the contributions of quark-line
disconnected diagrams which are very difficult to compute as they
involve propagators from all points on the lattice to themselves.
Consequently, they require ensembles of gauge configurations that are
an order of magnitude larger to achieve the same precision as in the
corresponding connected diagrams \cite{Gusken:1999te} even when
stochastic estimator techniques
\cite{Bitar:1988bb,Dong:1993pk,Eicker:1996gk,Thron:1997iy,Viehoff:1997wi,Gusken:1998wy}
(which improve greatly on the brute force approach) are applied.  In
the background field approach, disconnected contributions are included
automatically; for these calculations one requires a few additional,
moderately sized ensembles of configurations generated with modified
actions and this may be a more efficient means of calculating flavour
singlet quantities. In terms of statistics and identification of the
ground state, the external field method is somewhat similar to the
plateau accumulation method \cite{Gusken:1998wy}, but with the added
advantage of automatically including quark-line disconnected diagrams.
To extract physical quantities in either approach, calculations must
be repeated at different quark masses, volumes and lattice spacings
and then the necessary extrapolations must be performed. This may be
more difficult using external fields than in the operator insertion
approach where multi-mass techniques \cite{Frommer:1995ik} are well
developed (though similar techniques may also make the background
field calculations easier).  Numerical work to investigate the
relative costs of the different approaches is encouraged.

{\it Two baryons:} In the two baryon sector, exponentiation of
operators will also prove useful. It has recently been proposed that
background electroweak fields can be used to measure the electroweak
properties of two-nucleon states such as the deuteron magnetic moments
and the cross-section for neutrino breakup of the deuteron
\cite{Detmold:2004qn}. For such two-hadron systems, even a calculation
of the lowest energy levels involves large numbers of distinct Wick
contractions and is only presently becoming feasible
\cite{Fukugita:1994ve}; calculating matrix elements of such systems by
an operator insertion is probably beyond the limits of current
computational power.  However, by measuring finite volume shifts in
the low-lying two-particle energy levels in appropriate background
fields, the various electroweak properties can be determined using
effective field theory.  Again, the exponentiated operator approach is
especially suitable for flavour singlet quantities as disconnected
diagrams, which would be required in the operator insertion approach,
are included automatically.  Provided one can construct deuteron
interpolating fields that sufficiently project onto its different
polarisations, similar background field methods will yield the
quadrupole moment of the deuteron.

Another important phenomenon which background fields may prove useful
in investigating is the modification of structure functions in nuclei
-- the EMC effect \cite{Norton:2003cb}. The difference of the ratio
$F_2^d(x)/\left[F_2^p(x)+F_2^n(x)\right]$ from unity is the simplest
quantity in this regard (though experimentally the magnitude of the
effect increases with atomic number).  Important matrix elements that
one might consider to investigate this on the lattice are $\langle N\,
N|{\cal O}^{\mu_0\ldots\mu_n}_{\gamma,u+d}|N\, N\rangle$
\cite{EMCpaper} since these determine the structure functions of
two-nucleon states such as the deuteron via the operator product
expansion.  In a similar manner to the deuteron magnetic moment and
neutrino-deuteron breakup cross-sections \cite{Detmold:2004qn}, one
can exponentiate the twist-two operators to create background fields.
Then evaluating two-particle energy levels in these field and matching
to effective field theory will enable calculation of the required
matrix element.  Again, since these matrix elements are isoscalar, the
external field approach should be competitive with operator insertion
techniques\footnote{Indeed, the cost comparison becomes more
  favourable as the number of connected quark lines increases.}.
Finally, we note that the same sets of generalised background field
gauge configurations that one might use to calculate moments of
twist-two operators in the single baryon and meson sectors would also
be appropriate here.

In summary, background field methods in which an operator is
exponentiated into the action are powerful tools for evaluating
general hadronic matrix elements in lattice QCD. Here, we have
highlighted some of the features of this approach. The method is not
restricted to operators whose background fields have classical
analogues; for example, twist-two operators, such as that determining
the total angular momentum content of the proton, are ideal
candidates.  For flavour singlet matrix elements, disconnected
diagrams are automatically calculated in the external field procedure,
potentially leading to increased statistical precision in these
quantities.  The technique is also not limited to forward matrix
elements; objects such as the electromagnetic form-factors and
generalised parton distributions can also be computed with background
fields. These advantages come at the cost of requiring additional
ensembles of gauge configurations to be generated.

The author is grateful for discussions with C.-J.~D.~Lin,
W.~Melnitchouk, M.~J.~Savage, A.~Shindler and J.~M.~Zanotti. This work
is supported by the US Department of Energy under contract
DE-FG03-97ER41014.

\end{document}